\documentclass[12pt]{article}
\usepackage{epsfig,amsmath,graphics}
\usepackage{graphics,graphicx}
\usepackage{tabularx}
\usepackage{amsfonts}
\usepackage{amsmath}

\def\ds{\displaystyle}

\newcommand{\beq}{\begin{equation}}
\newcommand{\eeq}{\end{equation}}
\newcommand{\lb}{\label}
\newcommand{\beqar}{\begin{eqnarray}}
\newcommand{\eeqar}{\end{eqnarray}}
\newcommand{\barr}{\begin{array}}
\newcommand{\earr}{\end{array}}

\newcommand{\derp}[2]{\ds{\frac {\partial #1}{\partial #2}}}

\def\scalp{\mbox{\boldmath$\, \cdot \, $}}

\def\XXint#1#2#3{{\setbox0=\hbox{$#1{#2#3}{\int}$}
     \vcenter{\hbox{$#2#3$}}\kern-.5\wd0}}

\def\c{{\circ}}

\def\bD{\mbox{\boldmath${\it D}$}}

\def\bE{\mbox{\boldmath${\it E}$}}
\def\be{\mbox{\boldmath${\it e}$}}
\def\bF{\mbox{\boldmath${\it F}$}}

\def\bI{\mbox{\boldmath${\it I}$}}

\def\bm{\mbox{\boldmath${\it m}$}}

\def\bn{\mbox{\boldmath${\it n}$}}
\def\b0{\mbox{\boldmath${\it 0}$}}

\def\bS{\mbox{\boldmath${\it S}$}}

\def\bx{\mbox{\boldmath${\it x}$}}

\def\Id{\mbox{\boldmath${\it I}$}}

\def\bepsilon{\mbox{\boldmath${\epsilon}$}}

\def\btau{\mbox{\boldmath${\tau}$}}

\def\tr{{\rm tr}}

\def\Div{{\rm Div}}

\def\Curl{{\rm Curl}}
\def\Grad{{\rm Grad}}

\def\diag{{\rm diag}}

\def\APL{{\em Appl. Phys. Lett.\ }}

\def\JMPS{{\em J. Mech. Phys. Solids\ }}

\def\PRSL{{\em Proc. R. Soc. Lond.\ }}

\def\salto#1#2{
[\mbox{\hspace{-#1em}}[#2]\mbox{\hspace{-#1em}}]}

\begin{document}

\title{Performance of soft dielectric laminated composites}

\author{
   Massimiliano Gei, Roberta Springhetti, Eliana Bortot
\\
\\
\small{\sl{Department of Civil, Environmental and Mechanical Engineering (DICAM)}},
\\
\small{\sl{University of Trento, Via Mesiano 77, I-38123 Trento, Italy.}} \\
\small{\sl{Email: massimiliano.gei@unitn.it; web-page: www.ing.unitn.it/$\sim$mgei.}}}

\maketitle

\begin{abstract}
This paper contains a thorough investigation of the performance of electrically activated layered soft dielectric composite actuators
under plane deformation. Noting that the activation can be induced controlling either the voltage or the surface charge,
the overall behaviour of the system is obtained via homogenization at large strains taking either
the macroscopic electric field or the macroscopic electric displacement field as independent electrical variable.
The performance of a two-phase composite actuator compared to that of the homogeneous case is highlighted for few boundary-value problems
and for different values of stiffness and permittivity ratios between constituents being significant for applications, where the soft matrix is reinforced by a relatively small volume fraction of a stiff and high-permittivity phase.
For charge-controlled devices, it is shown that some composite
layouts admit, on one hand, the occurrence of pull-in/snap-through instabilities that can be exploited to design release-actuated systems, on the other, the possibility of thickening at increasing surface charge density.
\end{abstract}


\vspace*{5mm}


\section{Introduction}
A challenging research topic in the field of Dielectric Elastomer (DE) devices \cite{pelrine_science2000,carpietal} is the design and realization of effective composite materials,
exhibiting improved actuation properties with respect to the classical homogeneous materials employed so far, namely acrylic elastomers and silicones
\cite{huangcomp2004,lihuang2004,gallone2010,opris2010,liuleng2010,kofod2011,kofod2012,skov2012}.
The goal achievement is still far as a consequence of the problems encountered in the control of the microstructure during the phase mixing. In general, despite an improved performance at low voltage, the heterogeneous material results to be weaker with respect to electric breakdown \cite{kofod2011,kofod2012}.

On the other hand, theoretical modelling of dielectric elastomer composites has been facing great advances in the last few years, showing
that, in the small strain setting, suitably designed heterogeneous materials potentially reach the prescribed actuation strain as effect of
a voltage much lower than the one required by homogeneous specimens \cite{Tian} (more than one order of magnitude lower).
In the more appropriate finite strain framework, the mathematical formulation of predicting models and the analysis of
soft dielectric composites were considered only in few papers: deBotton et al. \cite{gal_limor_mams07} provided the first analysis, briefly discussing  the behaviour of laminates; Bertoldi and Gei \cite{maxkatia2011} and Rudykh and deBotton \cite{Rudykh} investigated in detail the homogenization and the micro- and macroscopic
stability of layered composites adopting the electric displacement field as the independent electric variable. Recently,
Ponte Casta\~neda and Siboni \cite{ppcsiboni} developed a complete homogenization theory for soft dielectrics reinforced by
high-permittivity particles.

The paper presents a complete analysis of the performance of rank-1 layered dielectric composites deforming nonlinearly under plane strain
conditions, investigating the electromechanical response relevant to different boundary-value problems
for several values of stiffness and permittivity ratios between phases, being meaningful towards applications, where the soft matrix is reinforced
by a relatively small volume fraction of a stiff and high-permittivity material. This contribution extends that
by deBotton et al. \cite{gal_limor_mams07}, providing new insights on the macroscopic response of such composites in terms of
the different types of electrical stimulation.
As heterogeneous bodies invariably involve an internal microstructure, even though each phase behaves isotropically, the homogenized model
results in an anisotropic constitutive equation. If anisotropy affects the dielectric properties, namely the relationship between the
electric field and the polarization, the actuation induced in a capacitor-like device based on this material by controlling the voltage
--thus setting the electric field-- is different than that obtained assigning the electric charge on its two opposite sides \cite{Kaltenbrunner}
--thus setting the electric displacement field--, as the two vector fields are not aligned.
This leads to different behaviours of the same system under the two types of actuation and then to a different macroscopic
performance.

While, on one side, composite materials represent a promising way to improve the overall performance of soft dielectric actuators,
on the other, they can be exploited in order to promote unconventional behaviors not arising in homogeneous systems,
such as the occurrence of a snap-through effect along the electromechanical loading path, which turns out to be useful for the realization of release-actuated transducers, or the thickening
effect shown by some layouts under charge-controlled actuation.

The paper is organized in five sections. In Sect. 2, the homogenization theory at finite strain for rank-1 layered composites is recalled
and specialized to the  two types of actuation analysed. In Sect. 3, the boundary-value problems taken into account to assess the behaviour of the
heterogeneous devices are described, while in Sect. 4 the relevant results of our analysis are illustrated. Finally,
conclusions are drawn in Sect. 5.

\section{Electromechanics of laminated composites under plane strain conditions}

The soft dielectric under investigation is a (rank-1) layered actuator obtained by repeating a unit cell consisting of two incompressible dielectric phases,
a soft matrix with low dielectric constant and a stiff and high-permittivity phase,
respectively denoted by 'b' and 'a', with a generic lamination angle $\theta$. Assuming that the length scale of the microstructure
is very small compared to the thickness of the actuator, the macroscopic behaviour of the system can be determined on the basis of the homogenization theory
\cite{ppcsiboni,hill72,ogden74,gal2005lam}.
If $h^{0a}$ and $h^{0b}$  represent the relevant phase thicknesses in the reference, stress-free configuration $B^0$
(see Fig. \ref{geometry}), the volume fractions are given by $c^a=h^{0a}/\left(h^{0a}+h^{0b}\right)$ and $c^b=1-c^a$, respectively.

\vspace{-.1cm}
\begin{figure}[h]
  \begin{center}
\includegraphics[width= 12 cm]{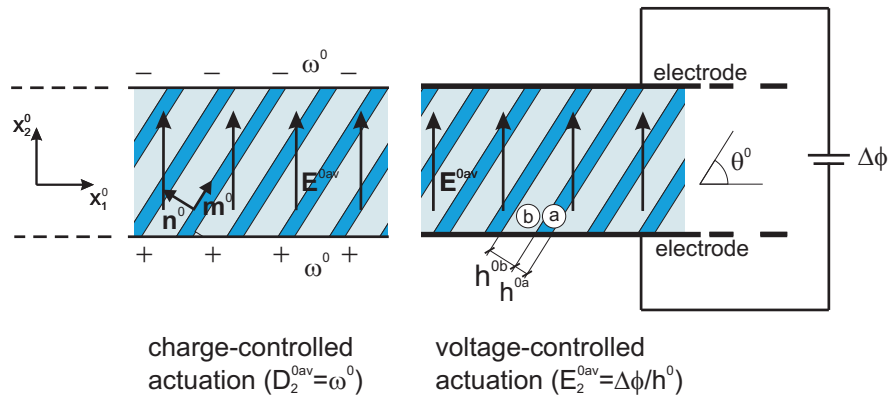}
\caption{\footnotesize Geometry of the reference configuration $B^0$ of a two-phase layered dielectric actuator subjected
to a voltage difference $\Delta \phi$ applied between electrodes (right) and to a surface charge distribution $\pm \omega^0$ spread on the two opposite boundaries of the system (left).}
    \lb{geometry}
 \end{center}
\end{figure}

The body is deformed by external electromechanical loadings reaching the current configuration $B$.
In a Lagrangian setting, with reference to the nonlinear theory of electroelasticity \cite{mcmeeking,dorf&ogde05acmc,suo,ask_ralf2013},
electromechanical equilibrium (in absence of body forces and free-volume charge)
of each phase along the deformation history, requires that\footnote{A standard notation for the variables introduced
in the paper is adopted. Symbols $\bx^0$ and $\bx$ represent points of $B^0$ and $B$, respectively, while differential
operators with capital letters are evaluated with reference to the variables in $B^0$. We denote by $\bS$ the total
first Piola-Kirchhoff stress, by $\bF=\Grad \bx$ the deformation gradient and by $\bD^0$ and $\bE^0$ the Lagrangian (or nominal) electric
displacement and the Lagrangian (or nominal) electric field, respectively.}
     \beq
     \lb{DivSDE}
     \Div\,\bS = \b0,\ \ \ \ \ \bS \bF^T=(\bS \bF^T)^T,\ \ \ \ \Div\, \bD^0=0,
    \ \ \ \ \Curl\, \bE^0=\b0,
     \eeq
and, since the electric field is conservative, $\bE^0=-\Grad\, \phi^0$, where $\phi^0(\bx^0)$ is the electrostatic potential.

Continuity at all the internal interfaces between phases \lq a' and \lq b' 
is enforced imposing the following jumps
\beq
\lb{jumpboundaryint}
\salto{0.1}{\bF}\bm^0=\b0,\ \ \ \ \salto{0.1}{\bS} \bn^0=\b0,\ \ \ \
\salto{0.1}{\bD^0}\scalp \bn^0=0,\ \ \ \ \bn^0 \times \salto{0.1}{\bE^0}=\b0,
\eeq
where $\salto{0.1}{f}=f^{a}-f^{b}$ and $\bn^0$ is the normal to the interface pointing toward 'b' that can be easily related to angle $\theta^0$.
Note that in plane strain, the last condition (continuity of the tangential component of the electric field) can be also written as
$\salto{0.1}{\bE^0}\scalp \bm^0=0$, where $\bm^0$ is aligned with the interface and such that $\bn^0 \scalp \bm^0=0$.

As standard in the electroelastic theory of continua at finite strain, a set of equations similar to (\ref{DivSDE}) can be
directly formulated in the current configuration $B$. In this case the total (true) stress $\btau$ replaces $\bS$, while the electric field $\bE$ and the electric
displacement $\bD$ take the place of $\bE^0$ and $\bD^0$, respectively, as the following transformations hold true:
\beq
\lb{lag_eul}
\btau=\bS \bF^T/J,\ \ \ \ \bD=\bF \bD^0/J,\ \ \ \ \bE=\bF^{-T}\bE^0,
\eeq
where $J=\det \bF=1$ as a consequence of incompressibility.

Here the heterogeneous actuator is conceived as a homogenized continuum, therefore its behaviour will be described on the basis of the macroscopic, \lq average',
quantities $\bS^{\rm av}$, $\bF^{\rm av}$, $\bD^{0\rm av}$ and $\bE^{0\rm av}$, for which the governing equations (\ref{DivSDE}) are still
valid in all the points of the body.

For the geometry considered (Fig. \ref{geometry}), the electric field outside the actuator vanishes and consequently, denoting by $\tilde\bn^0$
the outward unit normal to the boundaries, the boundary conditions take the expression
\beq
\lb{jumpext}
\bS^{\rm av} \tilde\bn^0=\b0,\ \ \ \
\bD^{0\rm av}\scalp \tilde \bn^0=-\omega^0, \ \ \ \ \tilde \bn^0 \times \salto{0.1}{\bE^{0\rm av}}=\b0.
\eeq


\subsection{Constitutive assumptions}
\lb{sec_cost_eq}

The constitutive equations provide the key relationships between the external electric input and the electromechanical response of the material.
It is assumed that {\em each phase} of the composite \emph{i)} is hyperelastic, obeying a neo-Hookean stress-strain law and \emph{ii)}
behaves as an ideal dielectric, i.e.
$\bD=\epsilon \bE$, where the dielectric constant (or permittivity) $\epsilon$ is independent of the deformation. On the other hand, as
both voltage actuation and charge actuation are investigated, two free energies,
both satisfying assumptions \emph{i)} and \emph{ii)}, are introduced: the first assumes $\bE^0$ as independent electric variable
[i.e. $W_{\rm e}=W_{\rm e}(\bF,\bE^0)$], while the second adopts the electric displacement $\bD^0$ [i.e. $W_{\rm d}=W_{\rm d}(\bF,\bD^0)$].
Therefore, the required expressions are
\beq
    \lb{neohookeanE}
    W_{\rm e}=\frac{\mu}{2}\left[\tr(\bF^T\bF)-3\right]-\frac{\epsilon}{2}(\bF^{(-T)} \bE^0\scalp \bF^{(-T)} \bE^0),
    \eeq
    \beq
    \lb{neohookeanD}
    W_{\rm d}=\frac{\mu}{2}\left[\tr(\bF^T\bF)-3\right]+\frac{1}{2\epsilon}(\bF \bD^0\scalp \bF \bD^0),
    \eeq
where $\mu$ is the shear modulus of the phase.
We remark that (\ref{neohookeanE}) and (\ref{neohookeanD}) are different expressions for the same energy, as can be verified using (\ref{lag_eul})
for incompressible materials and the proportionality between $\bE$ and $\bD$. The need of introducing both forms arises
in association with the different macroscopic behaviour induced by the two different actuation modalities taken into account. Thermodynamic
arguments provide the following general expressions for the constitutive laws
        \beq
        \lb{incomp_const_eq_1}
        \bS=\derp{W_{\rm e,d}}{\bF}-p \bF^{-T},\ \ \ \ \bD^0=-\derp{W_{\rm e}}{\bE^0},\ \ \ \  \bE^0=\derp{W_{\rm d}}{\bD^0},
        \eeq
where it is emphasized that the total stress is formally obtained in the same way for both approaches.
For the free energies (\ref{neohookeanE}) and (\ref{neohookeanD}) the explicit
form of (\ref{incomp_const_eq_1})$_1$ turns out to be
\beq
\lb{Siso_psi_neohookean}
\bS+p\,\bF^{-T}-\mu \bF=\frac{1}{\epsilon} \bF\bD^0 \otimes  \bD^0= \epsilon \bF^{-T} \bE^0 \otimes \bF^{-1} \bF^{-T} \bE^0,
\eeq
which corresponds to the same total stress
\beq
\lb{tau_neohookean}
\btau+p\bI-\mu \bF \bF^T=\frac{1}{\epsilon} \bD \otimes  \bD= \epsilon \bE \otimes \bE.
\eeq

Regardless of which formulation is adopted, the macroscopic free energy of the composite is obtained as the sum of the weighted
energies for each phase, namely
    $
    \lb{wav}
    W_{\rm e,d}^{\rm av}=c^{a}W_{\rm e,d}^{a}+c^{b}W_{\rm e,d}^{b}
    $
and the macroscopic total stress and electric field can be obtained from $W_{\rm e,d}^{\rm av}$ via constitutive equations
        \beq
        \lb{incomp_const_eq_comp}
        \bS^{\rm av}=\derp{W^{\rm av}_{\rm e,d}}{\bF^{\rm av}}-p^{\rm av}(\bF^{\rm av})^{-T},
        \ \ \ \  \bD^{0{\rm av}}=-\derp{W^{\rm av}_{\rm e}}{\bE^{0{\rm av}}},
        \ \ \ \  \bE^{0{\rm av}}=\derp{W^{\rm av}_{\rm d}}{\bD^{0{\rm av}}},
        \eeq
where, as indicated, all quantities with label \lq av' are meant to be average over a representative volume of the composite.

We point out that, on the basis of the homogenization procedure detailed below, an anisotropic macroscopic behaviour of the laminate is expected, both mechanically and electrically; the latter is evident in the Eulerian constitutive relationship
$\bD^{\rm av} = \bepsilon^{\rm av} \bE^{\rm av},$
where matrix $\bepsilon^{\rm av}$ depends on the properties of both phases, irrespective of the driving actuation.

\subsection{Homogenized solution controlling the voltage}
\lb{sec_hom_eq_voltage}

Under the assumption of a homogeneous response in each phase,
the macroscopic deformation gradient $\bF^{\rm av}$ and the average
Lagrangian electric displacement $\bE^{0\rm av}$ are the weighted sum of those in each phase \cite{gal_limor_mams07,gal2005lam}, namely
    \beq
    \lb{eqMacroE}
    \bF^{\rm av}=c^{a}\bF^{a}+c^{b}\bF^{b},\;\;\;\;
    \;\;\;\bE^{0\rm av}=c^{a} \bE^{0a}+c^{b} \bE^{0b}.
    \eeq
Interface compatibility (\ref{jumpboundaryint})$_1$, together with eq. (\ref{eqMacroE})$_1$ provide
    \beq
    \lb{fafbav}
    \bF^{a}=\bF^{\rm av}\left(\Id+\alpha\,c^{b}\bm^0\otimes\bn^0\right),\;\;\;\;
    \bF^{b}=\bF^{\rm av}\left(\Id-\alpha\,c^{a}\bm^0\otimes\bn^0\right),
    \eeq
where $\alpha$ is a real parameter. On the other hand, (\ref{jumpboundaryint})$_4$ requires that the jump in $\bE^0$ be along the
direction normal to the layers, namely
    \beq
    \lb{E2alt}
    \bE^{0a}-\bE^{0b}=\tilde \beta \bn^0,
    \eeq
where $\tilde\beta$ is another real parameter. It follows from (\ref{eqMacroE})$_2$ and (\ref{E2alt}) that
$$
    \bE^{0a}=\bE^{0\rm av}+c^b\tilde \beta \bn^0,\;\;\;\;\bE^{0b}=\bE^{0\rm av}-c^a\tilde\beta \bn^0.
    $$
Quantities $\alpha$ and $\tilde \beta$ are obtained enforcing (\ref{jumpboundaryint})$_{2,3}$ after the
substitution of the constitutive laws (\ref{incomp_const_eq_1})$_{1,3}$.
As both phases are described by the free energy (\ref{neohookeanE}), their expressions read:
    \beq
    \lb{alfabeta}
    \displaystyle{
    \alpha=\frac{\mu^{b}-\mu^{a}}{c^a \mu^{b}+c^b\mu^a}\frac{\bF^{\rm av}\bn^0\scalp\bF^{\rm av}\bm^0}
    {\bF^{\rm av}\bm^0\scalp\bF^{\rm av}\bm^0}},
     \eeq
\beq
\lb{betatilde}
 \tilde{\beta}=\frac{\epsilon^{b}-\epsilon^{a}}{c^b \epsilon^a+c^{a}\epsilon^{b}}
   \frac{(\bF^{\rm av})^{-T} \bE^{\rm 0 av} \scalp (\bF^{\rm av})^{-T}\bn^0}{{(\bF^{\rm av})^{-T}}\bn^0
   \scalp(\bF^{\rm av})^{-T}\bn^0}+\alpha \bE^{\rm 0 av}\scalp\bm^0.
\eeq
Condition (\ref{jumpboundaryint})$_{2}$
can be also used to evaluate the jump of the phase pressures $p^a$ and $p^b$, yielding
\beq
\label{pressurestenew}
\begin{split}
p^b-p^a=&\left\{\left[\left(\bF^{\text{av}}\right)^{-T}\bE^{0
\text{av}}\scalp\left(\bF^{\text{av}}\right)^{-T}\bn^0\right]^{2}
\frac{\epsilon^a \epsilon^b(\epsilon^a-\epsilon ^b)}
{(c^b\epsilon ^a+c^a\epsilon ^b)^2}\right.\\
&+ \mu ^{b }-\mu ^{a }\Bigg\}
\frac{1}{\left(\bF^{\text{av}}\right)^{-T}\bn^0\scalp\left(\bF^{\text{av}}\right)^{-T}\bn^0}.
\end{split}
\eeq

\subsection{Homogenized solution controlling the charge}
\lb{sec_hom_eq_charge}

A similar procedure can be followed for charge-controlled actuation, where the vector field $\bD^{\rm 0av}$ is now imposed. Assuming
    \beq
    \lb{eqMacroD}
    \bD^{0\rm av}=c^{a} \bD^{0a}+c^{b} \bD^{0b}
    \eeq
and (\ref{jumpboundaryint})$_3$, it turns out that
    \beq
    \lb{D2alter}
    \bD^{0a}-\bD^{0b}=\beta \bm^0,
    \eeq
where $\beta$ is a new real parameter. Employing again (\ref{fafbav}),
it follows from eqs. (\ref{eqMacroD})$_2$ and (\ref{D2alter}) that
    $$
    \bD^{0a}=\bD^{0\rm av}+c^b\beta \bm^0,\;\;\;\;\bD^{0b}=\bD^{0\rm av}-c^a\beta \bm^0.
    $$
Quantities $\alpha$ and $\beta$ are obtained enforcing (\ref{jumpboundaryint})$_{2,4}$ and, for the extended
neo-Hookean free energy (\ref{neohookeanD}), they take the expressions
    \beq
    \lb{alfabeta}
    \displaystyle{
    \alpha=\frac{\mu^{b}-\mu^{a}}{c^a \mu^{b}+c^b\mu^a}\frac{\bF^{\rm av}\bn^0\scalp\bF^{\rm av}\bm^0}
    {\bF^{\rm av}\bm^0\scalp\bF^{\rm av}\bm^0}},\ \ \ \
    \displaystyle{\beta=\frac{\epsilon^{a}-\epsilon^{b}}{c^a \epsilon^a+c^{b}\epsilon^{b}}
    \frac{\bF^{\rm av} \bD^{\rm 0 av}\scalp\bF^{\rm av}\bm^0}{\bF^{\rm av}\bm^0\scalp\bF^{\rm av}\bm^0}-
    \alpha \bD^{\rm 0 av}\scalp\bn^0},
     \eeq
that are equivalent to those calculated in \cite{maxkatia2011,Rudykh}.
Within this approach, the pressure difference is
\beq
\label{pressurekatia}
p^b-p^a=\left[ \frac{\epsilon^a -\epsilon^b}{\epsilon^a \epsilon^b} \left( \bD^{\rm 0 av} \scalp \bn^0\right)^2 +\mu^b-\mu^a \right]
\frac{1}{(\bF^{\rm av})^{-T} \bn^0 \scalp (\bF^{\rm av})^{-T} \bn^0}.
\eeq

\section{The boundary-value problems analysed}

The behaviour of the heterogeneous  actuator sketched in Fig.  \ref{geometry} is investigated for three different
plane-strain boundary-value problems.
As typical actuators  mainly consist in very thin specimens, in all of them we disregard the edge effects concentrated along the boundary
of the electrodes or related to a lack of homogeneity of the surface charge distribution. This means that we assume
that the Lagrangian electric field $\bE^{\rm 0 av}$ is directed along $x^0_2$.
Moreover, in all the following problems, we fix the rigid-body motion of the finite deformation such that the two straight boundaries remain
aligned with $x^0_1$ and the electric fields
$\bE^{\rm 0 av}$ and $\bE^{\rm av}=(\bF^{\rm av})^{-T}\bE^{\rm 0 av}$ act along the orthogonal direction, namely
\beq
\bE^{\rm 0 av}=E^{\rm 0 av} \be_2,\ \ \ \ \bE^{\rm av}=E^{\rm av} \be_2,
\eeq
where $\be_2$ is the unit vector associated with $x^0_2$.
The scalar relationship between $E^{\rm 0 av}$ and $E^{\rm av}$ depends on the macroscopic deformation gradient $\bF^{\rm av}$. In view of the
dielectric anisotropy, the electric displacements, $\bD^{\rm 0 av}$ and $\bD^{\rm av}$, are not in general aligned with $\be_2$.
Under voltage-controlled actuation, $E^{\rm 0 av}=\Delta \phi/h^0$, while when the charge is imposed,
$D^{\rm 0 av}_2=\omega^0$.

For all the problems treated the transverse direction remains traction free, namely
\beq
\lb{sav220}
S^{\rm av}_{22}=0,
\eeq
a condition that allows the determination of the unknown pressure $p^{\rm av}$.

In the first boundary-value problem (henceforth denoted by \lq bvp A'), we assume that the actuation will deform
the specimen macroscopically such that the principal strain directions correspond to $x^0_1$ and $x^0_2$ or, alternatively, that $\bF^{\rm av}$
admits the diagonal representation $\bF^{\rm av}=\diag[\lambda, 1/\lambda,1]$, where $\lambda$ is the longitudinal stretch.
The electromechanical response (i.e., the laws $\lambda$--$\Delta \phi$ or $\lambda$--$\omega^0$) can be determined
imposing $S^{\rm av}_{11}=0$ (null average longitudinal traction).
We note that the average shear stresses are not vanishing in this problem, nevertheless this deformation path is important for two reasons:
i) taking into account the slenderness of real devices, it can be concluded that shear deformability does not play a major role in the
estimation of the longitudinal stretch at moderate electrical excitation (as it will be clarified in the description of \lq bvp B');
ii) the deformation adopted here is usually employed to investigate micro and macroscopic instabilities of laminates at finite strain
\cite{gal_limor_mams07,maxkatia2011,Rudykh,gal2005lam}.

In the second problem (\lq bvp B') (see Fig. \ref{bvp}), the device will deform macroscopically at vanishing tractions, namely we impose, in addition
to eq. (\ref{sav220}), $S^{\rm av}_{11}=0$
and $S^{\rm av}_{12}=0$ along the electric activation. Due to the intrinsic macroscopic mechanical anisotropy, the admissible deformation
gradient is
\beq
\bF^{\rm av}=\left[
  \begin{array}{ccc}
    \lambda & \xi/\lambda & 0 \\
    0 & 1/\lambda & 0 \\
    0 & 0 & 1 \\
  \end{array}
\right],
\eeq
where $\xi$ is the amount of shear related to the shear angle $\gamma$ in the current configuration ($\tan \gamma=\xi$).
It is expected that at low voltage/charge the stretch $\lambda$ will be very close to that computed in \lq bvp A', while diverging at higher
electrical excitations.

In the third problem (\lq bvp C'), the composite membrane, of length $l^0$, is first prestretched (and clamped) longitudinally
at $\lambda=\lambda_{\rm pre}$
and then actuated applying the voltage or spraying the charge onto a limited portion of its boundaries (of length equal to $\lambda_{\rm pre} a^0$)
in the intermediate configuration. This will subdivide the body in a central \lq active' part and two lateral \lq passive' zones.
The current state of the actuator, namely the axial stretches inside the two portions, $\lambda^{\rm act}$ and $\lambda^{\rm pas}$, can be calculated
enforcing the invariancy of the total length and the continuity of the axial force at the interface between active and passive zones. This is
formalized by the system
\beq
\lambda_{\rm pre} l^0=\lambda^{\rm act} a^0+\lambda^{\rm pas} (l^0-a^0), \ \ \ \
S_{11}^{\rm av,act}=S_{11}^{\rm av,pas},
\eeq
where, upon substitution of the constitutive equations providing the expressions for $S_{11}^{\rm av,act}$,
$S_{11}^{\rm av,pas}$, the unknowns turn out to be $\lambda^{\rm act}$ and $\lambda^{\rm pas}$. This configuration was considered for homogeneous problems in
\cite{gei_bg_eap2011,puglisi_actpas} to analyse the properties of a periodic electrically-tunable bending-waveguide and in \cite{puglisi_actpas} to investigate
the stability of a prestretched DE actuator.

\begin{figure}[!tcb]
  \begin{center}
\includegraphics[width= 12 cm]{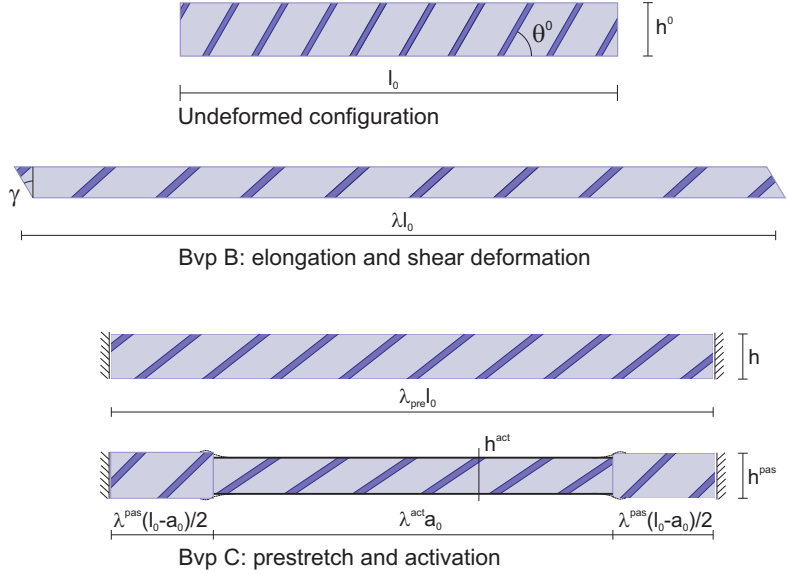}
\vspace{0.5 cm}
\caption{\footnotesize Analysed boundary-value problems for a layered composite actuator in the conditions sketched in Fig. \ref{geometry}. \lq Bvp A' is not reported.}
    \lb{bvp}
 \end{center}
\end{figure}

\section{Macroscopic performance}

\subsection{Voltage-controlled actuation}

\begin{figure}[!tcb]
  \begin{center}
\includegraphics[width= 7 cm]{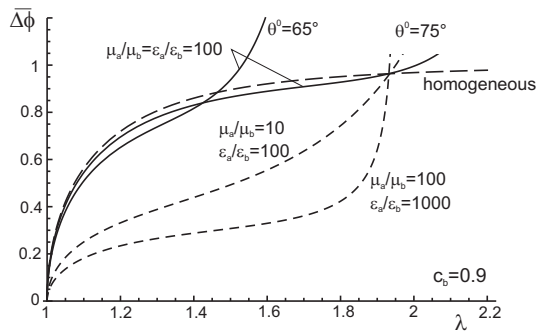}
\vspace{0.5 cm}
\caption{\footnotesize Performance of a voltage-controlled layered actuator (\lq bvp A') for different stiffness and permittivity
ratios and for two representative initial layer inclinations $\theta^0$
[$\overline{\Delta \phi}=(\Delta \phi/h^0)\sqrt{\epsilon^b/\mu^b}$, $\lambda$: longitudinal stretch].}
    \lb{e0av1}
 \end{center}
\end{figure}

The actuation behaviour of a device subjected to an increasing voltage $\Delta \phi$
is here investigated.  The properties of the composite are set by the volume fraction of the soft
matrix $c^b$, the angle of inclination of phases in the reference configuration $\theta^0$ and the ratios $m=\mu^a/\mu^b$ and
$r=\epsilon^a/\epsilon^b$. Results are provided in dimensionless form, so that the scale of voltage is given in terms of the quantity
$\overline{\Delta \phi}=(\Delta \phi/h^0)\sqrt{\epsilon^b/\mu^b}$.

Fig.  \ref{e0av1} refers to \lq bvp A', where the principal deformations are aligned with axes $x^0_1$ and $x^0_2$ introduced in Fig. \ref{geometry}.
The plots show that the performance enhancement of the heterogeneous actuator compared with the homogeneous case (where the device is exclusively composed of the soft polymer \lq b')
could be remarkable especially at low stretches
(i.e. the composite can reach the same stretch $\lambda$ at lower voltage).
The upper limit of the stretch range where this occurs  mainly depends on the
initial angle $\theta^0$, that is usually chosen
in the range $[60^\c, 90^\c]$ in order to exploit, along the deformation, the benefit ensuing from the finite rotation of the stiff phase.
When the current inclination reaches $\theta \approx 45^\c$ (simple geometrical considerations show that $\tan\theta=\tan\theta^0/\lambda^2$),
a strong stiffening in the actuation response takes place, as it becomes more and more difficult to  longitudinally stretch
the stiff phase beyond this limit. In the figure,
$\theta^0$ assumes the values of $\theta^0=65^\c$
and $75^\c$.
Even for the most favourable layer inclinations, when ratios $m$ and $r$ coincide, the performance improvement of the composite is very poor,
while it is definitely remarkable when $r$ is one order of magnitude higher than $m$.

In Fig. \ref{e0av1_scorrimento}, the deformation arising in \lq bvp B' is reported. As for the longitudinal stretch, the same considerations illustrated above
for \lq bvp A' still hold true, with an even worse performance of the composite in the case where $m$ and $r$ are coincident.
On the other hand, this increased longitudinal stiffness of the heterogeneous material is associated with the occurrence of shear deformations, which evolve very remarkably
with the voltage as shown in Fig. \ref{e0av1_scorrimento}b: beyond a sort of threshold in voltage, which depends on the ratios $m$ and $r$ of the two phases,
the shear angle exhibits large increments for small voltage increments, growing almost unboundedly.
It also appears that the lower the disparity of the two phases involved, the stiffer is the composite with respect to shearing, tending to the limit case of the
homogeneous material, for which $\gamma=0$, whatever voltage is applied.
In order to conceive actuators based on shear, it is also interesting to investigate the dependence of the attainable shear angles onto the fiber inclination, as pictured
in Fig. \ref{e0av1_scorrimento_angolo}. It appears that the behaviour of the composite is strongly influenced by its geometry and, also in this context, angles $\theta^0 \simeq 60^\c, 70^\c$
demonstrate to be the most advantageous.

\begin{figure}[!tcb]
  \begin{center}
\includegraphics[width= 8 cm]{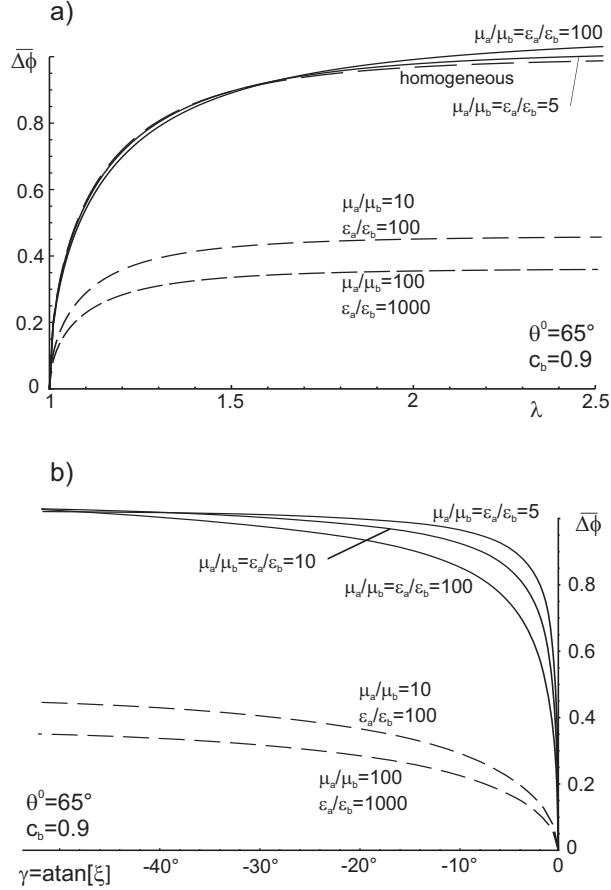}
\vspace{0.5 cm}
\caption{\footnotesize Performance of a voltage-controlled layered actuator  with vanishing normal and shear tractions
 (\lq bvp B') for different stiffness and permittivity
ratios ($\gamma$: shear angle). }
    \lb{e0av1_scorrimento}
 \end{center}
\end{figure}

\begin{figure}[!tcb]
  \begin{center}
\includegraphics[width= 8 cm]{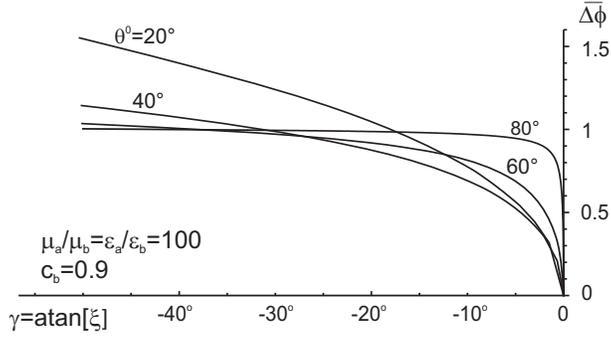}
\vspace{0.5 cm}
\caption{\footnotesize Performance of a voltage-controlled layered actuator: evolution of the shear angle $\gamma$ with the
layer inclination in the reference configuration $\theta^0$. }
    \lb{e0av1_scorrimento_angolo}
 \end{center}
\end{figure}

The actuation properties of the prestretched specimen described in \lq bvp C' are illustrated in Fig. \ref{actuator_eliana} where $c^b=0.9$, $a^0=0.3 l^0$
and the contrast parameters $m$ and $r$ both assume the representative value of 50. The specimen is prestretched at $\lambda_{\rm pre}=1.2$ (reaching a configuration
where $\theta=60.77^\c,\ 72.18^\c$ for $\theta^\c=65^\c, 75^\c$, respectively) and then actuated by application
of a voltage to the active part, that shrinks and elongates, while the remaining passive portion contracts longitudinally with a consequent decreasing of the axial tensile stress.
The comparison with the homogeneous case shows that for $60^\c<\theta^\c<85^\c$ the composite behaves more efficiently, displaying higher stretches $\lambda^{\rm act}$
up to the occurrence of the \lq loss of tension' of the passive part of the actuator.

\begin{figure}[!tcb]
  \begin{center}
\includegraphics[width= 8 cm]{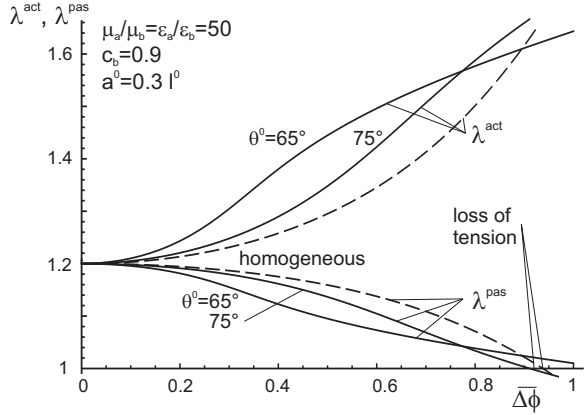}
\vspace{0.5 cm}
\caption{\footnotesize Performance of a prestretched, voltage-controlled layered actuator in the central, 'active' zone (\lq bvp C')
 for different stiffness and permittivity ratios ($\lambda_{\rm pre}=1.2$). }
    \lb{actuator_eliana}
 \end{center}
\end{figure}

\subsection{Charge-controlled actuation}

\begin{figure}[!htcb]
  \begin{center}
\includegraphics[width= 7 cm]{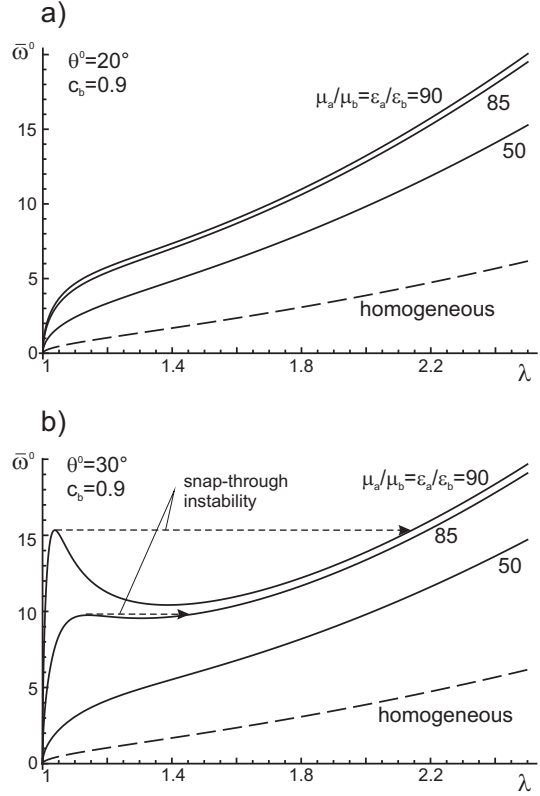}
\vspace{0.5 cm}
\caption{\footnotesize Performance of a bilayered, charge-controlled layered actuator (\lq bvp A'),
as a function of the layering angle and the contrast ratios $r=m$
between the two phases [$\bar \omega^0=\omega^0/\sqrt{\mu^b \epsilon^b}$, $\lambda$: longitudinal stretch].
a): $\theta^0=20^\c$, b): $\theta^0=30^\c$;
in part b) an electromechanical instability is induced for loading paths at high values of parameter $r$.}
    \lb{d0av}
 \end{center}
\end{figure}

Under charge-controlled actuation, we have never found an improvement of the macroscopic response in terms of stretch with respect to the homogeneous case. An example can be found
in Fig. \ref{d0av}a, where it is clear that the response of the composite (the axial stretch $\lambda$) at the same charge
($\bar \omega^0=\omega^0/\sqrt{\mu^b \epsilon^b}$)
is worse than that of the homogeneous actuator.
In any case, our aim is to show that such kind of actuation may induce interesting properties that can be exploited for the design of new type of actuators.

Comparing the plots a) and b) in Fig. \ref{d0av}, it can be noted that for some combinations of the constitutive parameters, the overall response is very sensitive to the
layering angle $\theta^0$, as plot b) is quite different from plot a), despite the small difference between the two angles (respectively, $\theta^0= 30^\c$ and $20^\c$). For values of the
contrast parameter higher than 80, the curves display a peak at low stretch, indicating the possibility of a snap-through instability triggered by pull-in. We highlight that a similar
behaviour is not possible for the homogeneous material (\ref{neohookeanD}) under plane deformations and therefore it represents a peculiar feature of the composite, that can be in principle
exploited to design released-actuated systems.

When shear deformation is not constrained, as for \lq bvp B', an overall behaviour as the one sketched in Fig. \ref{Dshear} has been found for $\theta^0 = 65^\c$,
where the dimensionless charge density has been plotted versus the shear angle for different values of the contrast parameters. In this context it
seems that in order to reach high shear angles,
while keeping the charge densites limited, the disparity of the two phases should be contained.

\begin{figure}[!h]
  \begin{center}
\includegraphics[width= 10 cm]{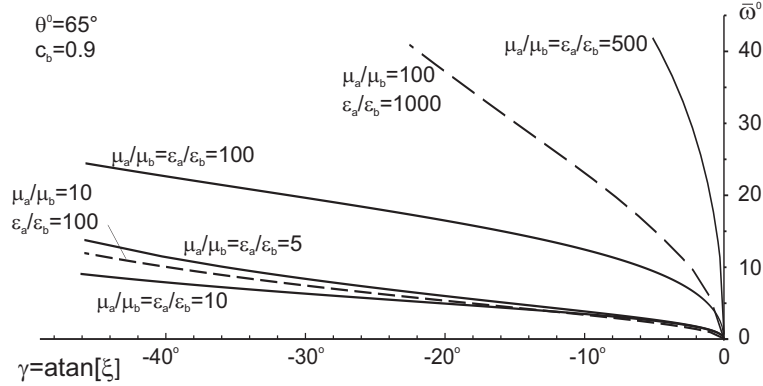}
\vspace{0.5 cm}
\caption{\footnotesize Performance of a charge-controlled layered actuator with vanishing normal and shear tractions
 (\lq bvp B') for different stiffness and permittivity
ratios.}
    \lb{Dshear}
 \end{center}
\end{figure}

Another interesting example is provided by the possibility of conceiving a system able to {\em increase} its thickness at an increasing surface charge. This is shown in Fig.
\ref{d0av_thick}, where for quite high-contrast parameters ($m=r=150$) and for the range $12.5^\c < \theta^0 < 42^\c$ the longitudinal stretch will diminish at different rate as a function
of $\bar \omega^0$.

\begin{figure}[!h]
  \begin{center}
\includegraphics[width= 7 cm]{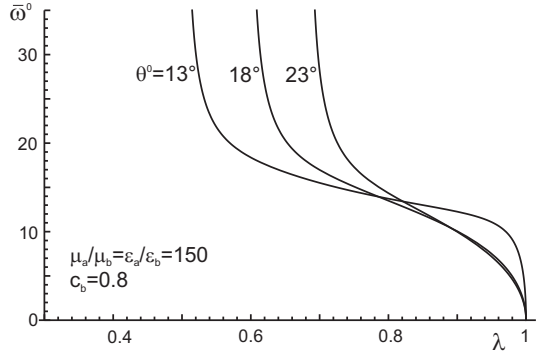}
\vspace{0.5 cm}
\caption{\footnotesize Thickening effect displayed by a charge-actuated layered composite (\lq bvp A').
For the parameters reported in the figure this
effect is displayed for an initial layer inclination in the range $12.5^\c < \theta^0 < 42^\c$.}
    \lb{d0av_thick}
 \end{center}
\end{figure}

\subsection{Overall plane strain states for traction-free actuators}

For a composite material undergoing plane strain deformation in the presence of a shear (\lq bvp B'), the overall deformation states which are
achievable on the basis of an electric actuation have
been collected in part a) of Fig. \ref{scorrimento_e0d0} for different values of the layer inclination $\theta^0$. Interestingly,
the deformation states only depend on the geometry and the constitutive
properties of phases, thus representing an intrinsic property of the composite material. In fact, no dependence has been highlighted on the way the
electric actuation is generated, as both voltage and charge control
lead to the same solution, which are indistinguishable in part a) of Fig. \ref{scorrimento_e0d0}. The part b) of the same figure pictures two possible deformed
configurations starting from the
same reference configuration, relevant to a material with layer inclination $\theta^0=65^\c$ and contrast ratios $m=r=100$. Each configuration can be reached through both a
voltage and a charge actuation for the indicated values of the controlled variable.

\begin{figure}[!tcb]
  \begin{center}
\includegraphics[width= 11 cm]{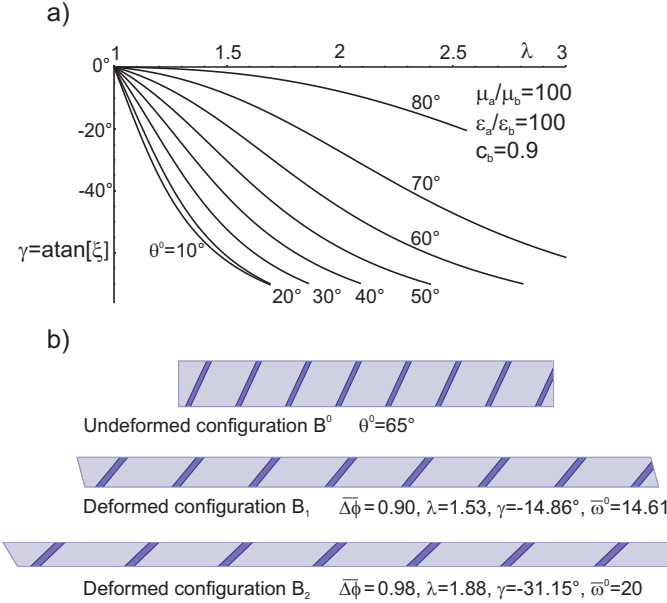}
\caption{\footnotesize Performance of a layered actuator with vanishing normal and shear tractions
 (\lq bvp B'). a) $\lambda$--$\gamma$ plot for different angles $\theta^0$, all deformation paths are valid for both types of actuation.
 b) undeformed and deformed configurations for a geometry with $\theta^0=65^\c$.}
    \lb{scorrimento_e0d0}
 \end{center}
\end{figure}

\section{Conclusions}

The overall electromechanical performance of rank-1 bi-layered soft dielectric composite actuators deforming in plane strain is analysed for different
boundary-value problems and for material parameters meaningful for applications. As mechanical and dielectric anisotropies are induced by the internal
microstructure, the actuation applying voltage provides, in general, a different response compared to that where surface
charge is controlled. For this reason, the homogenization problems where the electric displacement and the electric field are the independent
variables are both solved.

For voltage-controlled systems, the composite displays higher actuation strains than those of a homogeneous material
mainly for configurations allowing to exploit the finite rotations of the stiff and high-permittivity phase.
As an additional consequence of anisotropy, when the laminate is free to deform in the absence of shear stresses, a shear deformation appears developing
in a strongly nonlinear way, depending on the initial inclination of layers.
Interestingly, the deformation states of this mode depend only on the geometry and the constitutive
properties of phases, thus representing an intrinsic property of the composite material.

Under charge-controlled devices, the single-lamination layout proved not to be able to display enhanced actuation properties in terms of longitudinal stretch.
However, for some combinations of the material parameters, some peculiar phenomena may appear, such as
snap-through instabilities and thickening effects which can be usefully exploited in the realization of release-actuated transducers and thickness-mode devices, respectively.

\vspace{5 mm}

{\bf Acknowledgements}. The financial supports of PRIN grant no. 2009XWLFKW, financed by Italian Ministry of Education, University and Research,
and of the COST Action MP1003 \lq European Scientific Network for Artificial Muscles', financed by EU, are gratefully acknowledged.

\end{document}